\documentclass[letterpaper, 10 pt, conference]{ieeeconf}  

\IEEEoverridecommandlockouts
\usepackage{soul}

\usepackage{amsthm,amsmath,amssymb,amsfonts}
\usepackage{graphicx}
\usepackage{hyperref} 
\usepackage{textcomp}
\usepackage[x11names]{xcolor}
\usepackage{nicefrac} 
\usepackage{dsfont}
\usepackage{tikz} 
\usepackage{pgfplots}
\usepackage{pifont}
\usepackage{cleveref}
\usepackage[backend=bibtex, style=ieee]{biblatex}
\usepackage{setspace}
\usepackage{algpseudocode} 
\usepackage[]{algorithm}
\usepackage{fontawesome5}
\usepackage{subcaption}
\usepackage{float}
\usepackage{import}
\usepackage[export]{adjustbox}

\algrenewcommand\algorithmicindent{1.0em}%

\usepackage{bm}
\hypersetup{colorlinks=true,linkcolor=DodgerBlue3,filecolor=mnodea,urlcolor=cyan,citecolor=DodgerBlue3}

\usetikzlibrary{math,spy,arrows,arrows.meta, backgrounds,chains,positioning,shapes.geometric, shapes.symbols,shapes.arrows,patterns,automata, petri,topaths,plotmarks,decorations.fractals,pgfplots.groupplots}%
\pgfplotsset{compat=newest}

\usepgfplotslibrary{fillbetween,colorbrewer}

\newcommand{\grvect}[1]{\boldsymbol{#1}} 
\newcommand{\set}[1]{\mathcal{#1}} 
\newcommand{\ie}{\textit{i.e.,~}} 
\newcommand{\inneighbor}[1]{\set{N}_{#1}^{\footnotesize\texttt{in}}}

\newcommand{\outneighbor}[1]{\set{N}_{#1}^{\footnotesize\texttt{out}}}

\newcommand{\indegree}[1]{d_{#1}^{\footnotesize\texttt{in}}}
\newcommand{\outdegree}[1]{d_{#1}^{\footnotesize\texttt{out}}}
\newcommand{\diag}{\mathop{\mathrm{diag}}}

\DeclareTextFontCommand{\textbfit}{%
  \fontseries\bfdefault 
  \itshape
}

\newtheoremstyle{myTheoremStyle}
{}{}{\slshape}{}{\bfseries}{.}{ }{}\theoremstyle{myTheoremStyle}

\newtheorem{theorem}{Theorem}

\newtheorem{rem}{Remark}
\newtheorem{assumption}{Assumption}

\title{ Average Consensus over Directed Networks in Open Multi-Agent Systems with Acknowledgement Feedback }

 \author{
 Evagoras Makridis$^1$, Andreas Grammenos$^2$, Gabriele Oliva$^{3}$, Evangelia Kalyvianaki$^2$,\\ Christoforos N. Hadjicostis$^1$, and Themistoklis Charalambous$^{1}$
 \thanks{$^1$Department of Electrical and Computer Engineering, School of Engineering, University of Cyprus,
 2109 Aglantzia, Nicosia, Cyprus. E-mails: $\{$makridis.evagoras, hadjicostis.christoforos, charalambous.themistoklis$\}$@ucy.ac.cy.}
 \thanks{$^2$Department of Computer Science and Technology, University of Cambridge, Cambridge CB2 1TN, United Kingdom. E-mails: $\{$ag926, ek264$\}$@cl.cam.ac.uk.}
 \thanks{$^3$Department of Engineering, University Campus Bio-Medico of Rome, Via Alvaro del Portillo, 21--00128 Roma, Italy. E-mail: g.oliva@unicampus.it.}
 \thanks{The work of E. Makridis and T. Charalambous was partly funded by the project MINERVA, which received funding from the European Research Council (ERC) under the European Union's Horizon 2022 research and innovation programme (Grant Agreement No. 101044629).}
 }

\begin{document}

\maketitle
\begin{abstract}
In this paper, we address the distributed average consensus problem over directed networks in open multi-agent systems (OMAS), where the stability of the network is disrupted by frequent agent arrivals and departures, leading to a time-varying average consensus target. To tackle this challenge, we introduce a novel ratio consensus algorithm (\textsc{OpenRC}) based on acknowledgement feedback, designed to be robust to agent arrivals and departures, as well as to unbalanced directed network topologies. We demonstrate that when all active agents execute the \textsc{OpenRC} algorithm, the sum of their state variables remains constant during quiescent epochs when the network remains unchanged. By assuming eventual convergence during such quiescent periods following persistent variations in system composition and size, we prove the convergence of the \textsc{OpenRC} algorithm using column-stochasticity and mass-preservation properties. Finally, we apply and evaluate our proposed algorithm in a simulated environment, where agents are departing from and arriving in the network to highlight its resilience against changes in the network size and topology.
\end{abstract}

\begin{keywords}
distributed average consensus, open multi-agent systems, acknowledgement feedback.
\end{keywords}


\section{Introduction} 
In the domain of distributed and \emph{multi-agent systems} (MAS), a plethora of research has been conducted on collaborative decision-making and consensus. MAS consists of interconnected computing nodes, or agents, that exchange information with the goal of collaboratively attaining a shared understanding or decision on network-wide parameters, despite each relying solely on local information and data from neighboring agents. The lack of global knowledge about the states of other agents or network-wide parameters highlights the distributed nature of these systems, which are particularly suitable for tasks beyond the capabilities of individual master entities or those more efficiently managed by a cluster of simpler agents.

Notably, average consensus is a fundamental problem with broad applications across various fields. The goal of average consensus algorithms is to ensure that individual agent states or decisions converge to a common value, specifically the average of their initial states (the sum of the agents' initial states is often referred to as total mass). While significant advancements have been made in traditional MAS environments with a static, predefined number of agents, the introduction of \emph{Open Multi-Agent Systems} (OMAS) has added a new layer of complexity (refer to the recent thesis in \cite{de2022open} for an overview on the representations, limitations, and applications of OMAS). In OMAS settings, agents are allowed to dynamically join or leave the network during runtime, presenting challenges for maintaining stable consensus and effective collaboration.

Research efforts have spanned various aspects of OMAS dynamics, including trust and reputation models \cite{huynh2006integrated}, gossiping frameworks \cite{hendrickx2017open}, and the formation of short-term teams \cite{golpayegani2018participant,golpayegani2019using}. 
Further exploration into algorithms that handle predetermined periods of agent arrivals and departures \cite{7852357}, and strategies for managing time-varying network sizes \cite{de2020open}, alongside methods for estimating time-varying averages \cite{franceschelli2018proportional,franceschelli2020stability}, contribute to a rich body of knowledge aimed at enhancing the efficiency and reliability of MAS in a variety of applications.
Additionally, the application to distributed optimization~\cite{hendrickx2020stability,hayashi2022distributed} and learning~\cite{nakamura2023cooperative} showcase the depth of research in addressing the complexities introduced by agent dynamics.
Finally, it is worth mentioning the work in~\cite{oliva2023sum}, where the problem of OMAS interaction among agents with nonlinear coupling is addressed.

While prior studies that have explored various consensus frameworks within OMAS heavily rely on the assumption that the underlying graph is undirected, in this work we aim at designing an average consensus algorithm that can handle agent arrivals and departures over directed networks of time-varying size. In particular, we consider OMAS that consist of a set of agents that changes over time, within which agents can exchange information over a \emph{strongly connected}\footnote{A digraph is called strongly connected, if there exists a directed path between every pair of nodes in the graph.} and possibly unbalanced directed network. Within this setup, all active agents aim at iteratively computing the average of the initial values they had when they last entered the network. Each agent updates its state with local computation, which is often expressed as a linear combination of the state variables received by its immediate in-neighbors\footnote{The in-neighbors of a node are all the nodes in the network from which this node can directly receive information.}. 



Aiming towards the development of scalable average consensus algorithms that adapt to dynamic agent participation, our 
paper makes the following contributions.
\begin{itemize}
    \item 
    While previous works have primarily focused on undirected graphs, we introduce a fully distributed algorithm that can be executed by actively participating agents in OMAS over directed and possibly unbalanced graphs with one-bit acknowledgement feedback\footnote{Each acknowledgement feedback signal is sent over a narrowband error- and delay-free feedback channel. Hence, the network topology could still be assumed directed, although the feedback travels through undirected links.}. Our method eliminates the necessity of bidirectional information exchange to track a time-varying average consensus value, thus broadening its use in various applications.  
    \item Unlike classical approaches in undirected graphs, the proposed algorithm does not require individual agents to maintain the cumulative mass received from their in-neighbors. Instead, agents exploit the values they received from their in-neighbors, when they last joined the network, to ensure that the total mass of the active agents in the network remains invariant in between stable network conditions. Such an approach, not only reduces the computational burden on individual agents and enhances the scalability in large-scale systems, but also ensures that the active agents in the network can eventually reach the average consensus value in between stable network conditions.
\end{itemize}

\section{Preliminaries and Problem Setting}
\subsection{Open Network Model}\label{sec:open_network_model} 
In this work, we assume that agents are able to join and leave the network at will, provided that the network remains strongly connected at each time instant. In particular, we assume a finite set $\set{V}=\{v_1, \cdots, v_n\}$ that captures all $n$ agents potentially participating to the network.  
However, at each time step $k$, only a subset of the agents, denoted by $\set{V}(k) \subseteq  \set{V}$, is {\em active}, and we assume that interactions occur solely among active agents.
For analysis purposes, we denote the activation of the agents by a time-varying indicator vector $\grvect{\alpha}(k) \in \{ 0, 1 \}^n$, where $\alpha_j(k)=1$ if agent $v_j$ is active at time step $k$, \ie, if $v_j\in \mathcal{V}(k)$, while $\alpha_j(k)=0$, otherwise. Based on this activation vector, the interconnection topology of the active agents in the network is modeled by a (possibly unbalanced) time-varying digraph $\set{G}(k)=\{\set{V}(k), \set{E}(k)\}$. The number of agents active in the network at time $k$ is denoted by $n(k)=|\set{V}(k)|$. The interactions between active agents are captured by the set of digraph edges $\set{E}(k) \subseteq \set{V}(k) \times \set{V}(k)$. 
A directed edge denoted by $\varepsilon_{ji} \in \set{E}(k)$ indicates that node $v_j$ receives information from node $v_i$, \ie $v_i \rightarrow v_j$, at time step $k$. The set of all potential interactions is denoted by $\set{E}=\cup_{k=0}^{\infty} \set{E}(k)$. The nodes from which node $v_j$ receives information at time step $k$ are called in-neighbors of node $v_j$ at time step $k$, and belong to the set $\inneighbor{j}(k)=\{v_i \in \set{V}(k)| \varepsilon_{ji} \in \set{E}(k)\}$. The number of nodes in this in-neighborhood set is called the in-degree of node $v_j$ and is denoted by $\indegree{j}(k) = |\inneighbor{j}(k)|$. The nodes that receive information from node $v_j$ directly at time step $k$ are called out-neighbors of node $v_j$ at time step $k$, and belong to the set $\outneighbor{j}(k)=\{v_l \in \set{V}(k)| \varepsilon_{lj} \in \set{E}(k)\}$. The number of nodes in its out-neighborhood set of node $v_j$ at time step $k$ is called the out-degree at time step $k$ and is denoted by $\outdegree{j}(k)= |\outneighbor{j}(k)|$. The potential out-neighbors of node $v_j$ is given by $\outneighbor{j} = \cup_{k=0}^{\infty} \outneighbor{j}(k)$. Clearly, each node $v_j \in \set{V}(k)$ has immediate access to its own local state. 

The structure of the network at each time instant $k$ can be formally defined by the following three sets that distinguish the operating modes of the agents.

\textbf{Remaining:}
Agents that are currently active in the network at time $k$ and are not departing in the next step $k+1$, belong to the set of remaining agents
$$
\set{R}(k) = \set{V}(k) \cap \set{V}(k+1).
$$

\textbf{Arriving:} This set encompasses those agents that become active at time $k$ and is denoted by $\set{J}(k)$. In particular, we have that $\set{J}(k)$ is given by the agents that are active at time $k+1$, but are inactive at time $k$, i.e, 
$$
\set{J}(k) = \set{V}(k+1) \setminus \set{V}(k).
$$

\textbf{Departing:} This set collects those agents that leave the network at time $k$. The set is denoted by $\set{D}(k) $ and we have that
$$
\set{D}(k) = \set{V}(k) \setminus \set{V}(k+1).
$$

\begin{figure}[H]
  \begin{subfigure}[b]{0.5\linewidth}
    \centering
    \usetikzlibrary{shapes.geometric}
\usetikzlibrary{backgrounds}

\begin{tikzpicture}[>=stealth, thick,
   shorten >=3pt,
   shorten <=3pt,
   node distance=0.3cm,
   on grid,
   auto,
  ]
  
\tikzset{node distance = 1cm and 1cm}
\tikzset{invisible/.style={draw=none, thin, circle, fill=none, minimum size=0.05cm}}
\tikzset{main/.style={draw=black,thick, draw, circle, fill=gray!40, minimum size=0.4cm}}
\tikzset{label/.style={draw=none, circle, fill=none, minimum size=0.8cm}}
\tikzset{edge/.style ={draw=black, thick, ->,> = latex'}}
\tikzset{layer/.style={draw=black!60,minimum width=3cm,minimum height=2cm,shape=rectangle,densely dashed,fill=white}}

\node[layer] at (-0.65,-0.55) {};

\node[main] (v1) at (0.5,0) {}; 
\node[main,densely dashed,draw=gray!40,fill=gray!15] (vjoin) at (-1.8,-0.5) {}; 
\node[main] (v4) at (-0.7,-1.2) {};

\node[main] (vdepart) at (0.3,-1.1) {}; 

\node[main] (v5) at (-0.9,0) {};
\path[->] (v5) edge[bend right=10] node {} (v1);
\path[->] (v1) edge[bend right=10] node {} (v5);
\path[->] (v4) edge[bend left=10] node {} (v5);

\path[->] (vdepart) edge[bend left=10] node {} (v5);
\path[->] (vdepart) edge[bend left=10] node {} (v1);
\path[->] (v4) edge[bend left=10] node {} (vdepart);
\path[->] (v4) edge[bend left=10] node {} (v5);
\path[->] (vjoin) edge[bend left=15,dashed,gray!40] node {} (v4);
\path[->] (v1) edge[bend right=10] node {} (v4);
\path[->] (v4) edge[bend left=15,dashed,gray!40] node {} (vjoin);
\path[->] (vjoin) edge[bend left=15,dashed,gray!40] node {} (v5);

\end{tikzpicture}    \caption{Network $\set{V}(0) \subset \set{V}$} 
    \label{fig:a} 
    \vspace{2ex}
  \end{subfigure}
  \begin{subfigure}[b]{0.5\linewidth}
    \centering
    \usetikzlibrary{shapes.geometric}
\usetikzlibrary{backgrounds}

\begin{tikzpicture}[>=stealth, thick,
   shorten >=3pt,
   shorten <=3pt,
   node distance=0.3cm,
   on grid,
   auto,
  ]
  
\tikzset{node distance = 3cm and 2cm}
\tikzset{invisible/.style={draw=none, thin, circle, fill=none, minimum size=0.05cm}}
\tikzset{main/.style={draw=black, thick, draw, circle, fill=gray!40, minimum size=0.4cm}}
\tikzset{feedbackNodes/.style={draw=black, thick, draw, shape=ellipse, fill=gray!5, minimum size=0.6cm}}
\tikzset{label/.style={draw=none, circle, fill=none, minimum size=0.8cm}}
\tikzset{edge/.style ={draw=black, thick, ->,> = latex'}}
\tikzset{layer/.style={draw=black!60,minimum width=3cm,minimum height=2cm,shape=rectangle,densely dashed,fill=white}}

\node[layer] at (-0.65,-0.55) {};

\node[main] (v1) at (0.5,0) {}; 

\node[main,fill=SpringGreen4!30,inner sep=0,outer sep=0] (vjoin) at (-1.8,-0.5) {{\textcolor{SpringGreen4!80}{\small\faPlusCircle}}}; 

\node[main] (v4) at (-0.7,-1.2) {};

\node[main] (v5) at (-0.9,0) {};
\path[->] (v5) edge[bend right=10] node {} (v1);
\path[->] (v1) edge[bend right=10] node {} (v5);
\path[->] (v4) edge[bend left=10] node {} (v5);

\path[->] (vjoin) edge[bend left=15,draw=SpringGreen4!80] node {} (v4);
\path[->] (v1) edge[bend right=10] node {} (v4);
\path[->] (v4) edge[bend left=15,draw=SpringGreen4!80] node {} (vjoin);
\path[->] (vjoin) edge[bend left=15,draw=SpringGreen4!80] node {} (v5);

\node[main] (vdepart) at (0.3,-1.1) {}; 

\path[->] (vdepart) edge[bend left=10] node {} (v5);
\path[->] (vdepart) edge[bend left=10] node {} (v1);
\path[->] (v4) edge[bend left=10] node {} (vdepart);

\end{tikzpicture}    
    \caption{Agent arrival} 
    \label{fig:b} 
    \vspace{2ex}
  \end{subfigure} 
  \begin{subfigure}[b]{0.5\linewidth}
    \centering
    \usetikzlibrary{shapes.geometric}
\usetikzlibrary{backgrounds}

\begin{tikzpicture}[>=stealth, thick,
   shorten >=3pt,
   shorten <=3pt,
   node distance=0.3cm,
   on grid,
   auto,
  ]
  
\tikzset{node distance = 1cm and 1cm}
\tikzset{invisible/.style={draw=none, thin, circle, fill=none, minimum size=0.05cm}}
\tikzset{main/.style={draw=black, thick, draw, circle, fill=gray!40, minimum size=0.4cm}}
\tikzset{label/.style={draw=none, circle, fill=none, minimum size=0.8cm}}
\tikzset{edge/.style ={draw=black, thick, ->,> = latex'}}
\tikzset{layer/.style={draw=black!60,minimum width=3cm,minimum height=2cm,shape=rectangle,densely dashed,fill=white}}

\node[layer] at (-0.65,-0.55) {};

\node[main] (v1) at (0.5,0) {}; 
\node[main] (vjoin) at (-1.8,-0.5) {}; 
\node[main] (v4) at (-0.7,-1.2) {};

\node[main,fill=Coral3!30,inner sep=0,outer sep=0] (vdepart) at (0.3,-1.1) {{\textcolor{Coral3!80}{\small\faMinusCircle}}}; 

\node[main] (v5) at (-0.9,0) {};
\path[->] (v5) edge[bend right=10] node {} (v1);
\path[->] (v1) edge[bend right=10] node {} (v5);
\path[->] (v4) edge[bend left=10] node {} (v5);

\path[->] (vdepart) edge[bend left=10,draw=Coral3!80] node {} (v5);
\path[->] (vdepart) edge[bend left=10,draw=Coral3!80] node {} (v1);
\path[->] (v4) edge[bend left=10,draw=Coral3!80] node {} (vdepart);
\path[->] (v4) edge[bend left=10] node {} (v5);
\path[->] (vjoin) edge[bend left=15] node {} (v4);
\path[->] (v1) edge[bend right=10] node {} (v4);
\path[->] (v4) edge[bend left=15] node {} (vjoin);
\path[->] (vjoin) edge[bend left=15] node {} (v5);

\end{tikzpicture}
    \caption{Agent departure} 
    \label{fig:c} 
  \end{subfigure}
  \begin{subfigure}[b]{0.5\linewidth}
    \centering
    \usetikzlibrary{shapes.geometric}
\usetikzlibrary{backgrounds}

\begin{tikzpicture}[>=stealth, thick,
   shorten >=3pt,
   shorten <=3pt,
   node distance=0.3cm,
   on grid,
   auto,
  ]
  
\tikzset{node distance = 1cm and 1cm}
\tikzset{invisible/.style={draw=none, thin, circle, fill=none, minimum size=0.05cm}}
\tikzset{main/.style={draw=black, thick, draw, circle, fill=gray!40, minimum size=0.4cm}}
\tikzset{label/.style={draw=none, circle, fill=none, minimum size=0.8cm}}
\tikzset{edge/.style ={draw=black, thick, ->,> = latex'}}
\tikzset{layer/.style={draw=black!60,minimum width=3cm,minimum height=2cm,shape=rectangle,densely dashed,fill=white}}

\node[layer] at (-0.65,-0.55) {};

\node[main] (v1) at (0.5,0) {}; 
\node[main] (vjoin) at (-1.8,-0.5) {}; 
\node[main] (v4) at (-0.7,-1.2) {};

\node[main] (v5) at (-0.9,0) {};
\path[->] (v5) edge[bend right=10] node {} (v1);
\path[->] (v1) edge[bend right=10] node {} (v5);
\path[->] (v4) edge[bend left=10] node {} (v5);

\path[->] (v4) edge[bend left=10] node {} (v5);
\path[->] (vjoin) edge[bend left=15] node {} (v4);
\path[->] (v1) edge[bend right=10] node {} (v4);
\path[->] (v4) edge[bend left=15] node {} (vjoin);
\path[->] (vjoin) edge[bend left=15] node {} (v5);

\node[main,fill=gray!15,draw=gray!40,densely dashed] (vdepart) at (0.3,-1.1) {}; 
\path[->] (vdepart) edge[bend left=10,draw=gray!40,dashed] node {} (v5);
\path[->] (vdepart) edge[bend left=10,draw=gray!40,dashed] node {} (v1);
\path[->] (v4) edge[bend left=10,draw=gray!40,dashed] node {} (vdepart);
\end{tikzpicture}    
    \caption{Stable network} 
    \label{fig:d} 
  \end{subfigure} 
  \caption{OMAS with directional information flow. The size of the network changes as time progresses with agents joining the network, and present ones departing from the network.}
  \label{fig:omas_diagram}
\end{figure}
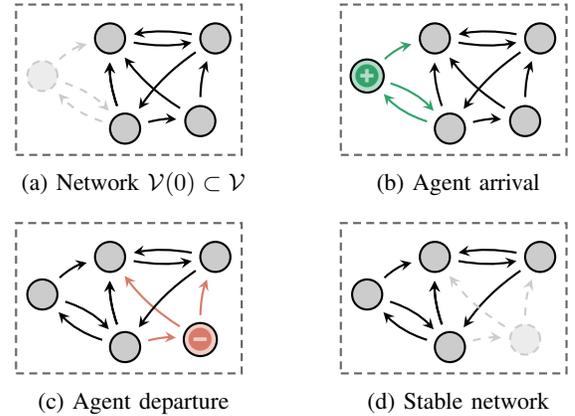

Fig.~\ref{fig:omas_diagram} depicts shapshots of the network considered in an OMAS framework, at different time instants.

\subsection{Assumptions}
In order to present the proposed problem statement, let us now state some assumptions.

\begin{assumption}[Graph strong connectivity]
    \label{ass:1}
    The directed graph $\set{G}(k)=\big\{\set{V}(k),\set{E}(k)\big\}$ is strongly-connected at all time instants $k \geq 0$.
\end{assumption}
\begin{assumption}[Acknowledgement feedback]
    \label{ass:2}
    For all times $k$ and each link $\varepsilon_{ji} \in \set{E}(k)$, we assume that agent $v_j$ is able to transmit a one-bit acknowledgement message to $v_i$. 
\end{assumption}
Assumption~\ref{ass:1} is common practice in the OMAS literature (connected for undirected graphs) \cite{dashti2022distributed}, and, as discussed later in the paper, guarantees that the sum of the states of the active agents equals the sum of their initial states. As we will discuss later on, from a time instant $k^{\prime}$ when the network composition stops evolving (no agent departures/arrivals occur) one can relax the strong connectivity assumption on the digraph $\set{G}(k)$ at time steps $k\geq k^{\prime}$, by only requiring a sequence of jointly strongly connected graphs (refer to \cite[Remark 3.4]{hadjicostis2018distributed} for more information).

Assumption~\ref{ass:2} is motivated by acknowledgement messages used by various protocols such as TCP, ARQ/HARQ, and ALOHA. 
Acknowledgement messages are utilized to confirm receipt of information and combat channel errors ensuring reliable transmissions over unreliable channels. These messages are assumed to be narrowband signals communicated over feedback channels, thus, their  presence does not necessarily conflict with the directional data channel. In the context of average consensus, acknowledgement messages are often used to ensure convergence to the exact average consensus value in error-prone networks, and allow the agents to determine their out-degree at each time instant~\cite{makridis2023utilizing}.


\subsection{Problem Statement}
Consider an open network as modeled in \S\ref{sec:open_network_model}. The goal of the active agents $v_j \in \set{V}(k)$ at time $k$ is to collaboratively compute the time-varying average of their \emph{joining masses}, \ie $\widehat{x}_j(k)$, albeit the changes in the composition and size of the network $\set{V}(k)$. In particular, the target time-varying average consensus value of the active agents is given by
\begin{align}\label{prob:1}
    \bar{x}(k)=\frac{1}{|\set{V}(k)|}\sum_{v_j\in \set{V}(k)} \widehat{x}_j(k),
\end{align}
where $\widehat{x}_j(k)$ tracks the joining mass, \ie the mass of agent $v_j$ when it last joins the network, defined as:
\begin{align}
    \!\!\widehat{x}_j(k) &=
    \begin{cases}
        m_j(\max\{r\in\set{K}_j\mid r \leq k\}), & \text{ if } \alpha_j(k) = 1,\\ 
        0, & \text{ otherwise, } 
    \end{cases}
\end{align} 
where $\set{K}_j$ is the set of times when agent $v_j$ (re)-joins the network, and $m_{j}(r) \in \mathbb{R}$ denote the \emph{joining mass}, with which agent $v_j$ enters the network at time $r \in \set{K}_j$ (clearly, for $r\in\set{K}_j$, we have $\alpha_j(r)=1$ and $\alpha_j(r-1)=0$). In other words, $\widehat{x}_j(k)$ is updated only when agent $v_j$ becomes active and it remains constant and equal to the joining mass value with which it enters the network at the latest (largest) $r \in \mathcal{K}_j$ with $r\leq k$. When $v_j \notin \set{V}(k)$, we can take $\widehat{x}_j(k)=0$.

Clearly, asymptotic convergence in such a setting cannot be achieved since, with each departure or arrival of agents, a perturbation on the time-varying average consensus value is generated. Here, it is important to note that, in gossip-based algorithms, even when agents may approach some steady state behavior temporarily (between consecutive arrival/departure events), the average consensus value $\bar{x}(k)$ cannot be reached exactly. This is due to the incapability of such algorithms to eliminate instantaneously the outdated information about agents that depart from the network, as well as the absence of a global variable that tracks the current number of active agents $n(k)$ in the network.

In this work, however, we propose a fully distributed algorithm, hereinafter referred to as \textsc{OpenRC}, which is capable to instantaneously eliminate outdated information from agents that leave the network. Therefore, active agents drive their states towards the perturbed, yet correct, average consensus value (of the currently active agents), by utilizing the joining and departing mass of the agents departing the network. This information is broadcasted over directional data channels prior to an agent's departure, to its out-neighbors, with an appropriate weight that is assigned by exploiting the corresponding acknowledgement signals.

\section{\textsc{OpenRC} Algorithm}

To compute the time-varying average consensus value in \eqref{prob:1}, we follow the ratio consensus paradigm in which each agent $v_j$ maintains two variables ${x}_j(k)$ and $y_j(k)$ (initialized upon activation at a given time $r$ as $\widehat{x}_j(r)$ and to $\widehat{y}_j(r)=1$, respectively). Here it is important to note that, $y_j(k)$ is an auxiliary variable which aims at compensating for the directional information flow (for more information, we refer the readers to \cite{hadjicostis2018distributed}). 
First, suppose that agent $v_j$ sends information to its out-neighbors $v_l \in \outneighbor{j}(k)$ at iteration $k$. We assume that, at the beginning of each round $k$, the out-neighbors of $v_j$ are able to send an acknowledgement feedback signal $f_{jl}(k)$ to $v_j$. In particular, for all potential out-neighbors $v_l\in\outneighbor{j}$, we have
\begin{align}
    f_{jl}(k) =
    \begin{cases}
        1, & \text{ if } v_l\in\set{R}(k) \land v_j\in\set{V}(k),\\
        0, & \text{ if } v_l\in\set{D}(k).
    \end{cases}
\end{align}
Based on the received feedback from each out-neighbor $v_l\in\outneighbor{j}$, encoded by $f_{jl}(k)$, agent $v_j$ is able to compute its currently active out-neighbors at each time $k$ by
$$
|\outneighbor{j}(k)\cap\mathcal{R}(k)|=\sum_{v_l \in \outneighbor{j}(k)} f_{jl}(k).
$$
In turn, the above quantities allow each agent $v_j$ to assign weights on its out-going links that scale the values to be sent to its currently active out-neighbors. Appropriate scaling of the out-going information is crucial to ensure that the total mass of the active agents in the network remains invariant in between stable network conditions. In what follows, we design the weight assignment from the perspective of node $v_j$, for its out-going links $\varepsilon_{lj} \in \set{E}(k)$.

\noindent\textbf{Arriving mode:}
When agent $v_j$ arrives in the network, \ie $v_j\in \mathcal{J}(k)$, it simply registers its joining mass by setting\begin{subequations}\label{eq:rc-omasJ}
\begin{align}
    x_j(k+1) &= \widehat{x}_j(k+1)
    \label{eq:rc-omasJ:1}\\
    y_j(k+1) &= \widehat{y}_j(k+1)
    \label{eq:rc-omasJ:2}\\
    z_j(k+1) &= {x_j(k+1)}/{y_j(k+1)},\label{eq:rc-omasJ:3}
\end{align}
and starts interacting with active neighboring agents at the next time step.
\end{subequations}

\begin{figure}[h]
    \centering
    \includegraphics[width=0.75\linewidth]{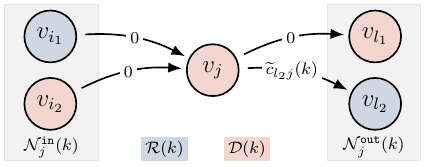}
    \caption{Weight assignment of node $v_j \in \set{D}(k)$. Nodes denoted in light blue belong in $\set{R}(k)$, while the ones denoted in light red belong in $\set{D}(k)$.}
    \label{fig:openrc_weights_departing}
\end{figure}

\noindent\textbf{Departing mode:} When agent $v_j$ departs from the network, \ie $v_j\in \set{D}(k)$, it broadcasts $\widetilde{c}_{lj}(k)(x_j(k)-\widehat{x}_j(k))$ and $\widetilde{c}_{lj}(k)(y_j(k)-\widehat{y}_j(k))$ to the out-neighbors $v_l \in \outneighbor{j}(k) \setminus  \set{D}(k)$ that remain in the network, as shown in Fig.~\ref{fig:openrc_weights_departing}, with the weights $\widetilde{c}_{lj}(k)$ given by
\begin{equation}
\begin{aligned}\label{eq:c-weightstilde}
    \widetilde{c}_{lj}(k) &= \dfrac{a_j(k)(1-a_j(k+1))f_{jl}(k)}{\sum_{v_l\in \outneighbor{j}(k)} f_{jl}(k)}\\
    &=\begin{cases}
    \dfrac{1}{|\outneighbor{j}(k)\cap\mathcal{R}(k)|},\!\!\!& \text{ if } v_l \in \outneighbor{j}(k)\setminus\mathcal{D}(k)\\
     0, &\text{ otherwise}.
    \end{cases}
\end{aligned}
\end{equation}

\noindent Note that $\tilde{c}_{lj}=0$ if $v_j \notin \set{D}(k)$. After a departing agent $v_j \in \set{D}(k)$ broadcasts its values, it departs from the network and hence we can consider that $x_j(k+1) = 0$, $y_j(k+1) = 0$, and $z_j(k+1) = 0$.

\begin{figure}[h]
    \centering
    \includegraphics[width=0.75\linewidth]{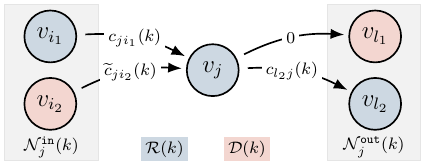}
    \caption{Weight assignment of node $v_j \in \set{R}(k)$. Nodes denoted in light blue belong in $\set{R}(k)$, while the ones denoted in light red belong in $\set{D}(k)$.}
    \label{fig:openrc_weights_remaining}
\end{figure}

\noindent\textbf{Remaining mode:} When agent $v_j$ remains in the network, \ie $v_j\in \set{R}(k)$, it broadcasts $c_{lj}(k)x_j(k)$ and $c_{lj}(k) y_j(k)$ to its out-neighbors $v_l \in \outneighbor{j}(k)\cap \set{R}(k)$, as shown in Fig.~\ref{fig:openrc_weights_remaining}, with the weights $c_{lj}(k)$ given by
\begin{equation}
\begin{aligned}\label{eq:c-weights}
    c_{lj}(k)&=\dfrac{a_j(k)a_j(k+1)f_{jl}(k)}{1+\sum_{v_l\in \outneighbor{j}(k)} f_{jl}(k)}\\
    &=\begin{cases}
    \dfrac{1}{1+|\outneighbor{j}(k)\cap\mathcal{R}(k)|},\!\!\!& \text{ if } v_l\in\mathcal{M}_j(k),\\
     0, &\text{ otherwise},
    \end{cases}
\end{aligned}
\end{equation}
where $\mathcal{M}_j(k)=\{\outneighbor{j}(k) \cap \mathcal{R}(k)\}\cup\{v_j\}$. Note that $c_{lj}(k)=0$ if $v_j \notin \set{R}(k)$.

In general, at time step $k$, each active agent $v_j \in \mathcal{V}(k)$ broadcasts the values
\begin{align}
    \label{eq:zetax}
    \zeta^{(x)}_j(k) &= c_{lj}(k)x_j(k)+\widetilde{c}_{lj}(k)(x_j(k)-\widehat{x}_j(k)),\\
    \label{eq:zetay}
    \zeta^{(y)}_j(k) &= c_{lj}(k)y_j(k)+\widetilde{c}_{lj}(k)(y_j(k)-\widehat{y}_j(k)),
\end{align}
to all its currently active out-neighbors. At the same time, upon the reception of information arrived within time slot $k$ from the in-neighbors of $v_j$, each agent $v_j\in \mathcal{R}(k)$ updates its local variables as follows:
{\small
\begin{subequations}\label{eq:rc-omas}
\begin{align}
    \!x_j(k+1) &= \!\!\!\! \sum_{v_i\in \mathcal{I}^{\mathcal{R}}_j\!(k) } \!\!\!\! c_{ji}(k) x_i(k) + \!\!\!\! \sum_{v_i \in \mathcal{I}^{\mathcal{D}}_j\!(k)} \!\!\!\! \widetilde{c}_{ji}(k) \big( x_i(k)-\widehat{x}_i(k) \big) \label{eq:rc-omas:1},\\
    \!y_j(k+1) &= \!\!\!\! \sum_{v_i\in \mathcal{I}^{\mathcal{R}}_j\!(k) } \!\!\!\! c_{ji}(k) y_i(k) + \!\!\!\! \sum_{v_i \in \mathcal{I}^{\mathcal{D}}_j\!(k)} \!\!\!\! \widetilde{c}_{ji}(k) \big( y_i(k)-\widehat{y}_i(k) \big) \label{eq:rc-omas:2},\\
    \!z_j(k+1) &= {x_j(k+1)}/{y_j(k+1)}.\label{eq:rc-omas:3}
\end{align}
\end{subequations}}%
where $\mathcal{I}^{\mathcal{R}}_j(k) = \{ \inneighbor{j}(k) \cap \set{R}(k) \} \cup \{ v_j \}$ and $\mathcal{I}^{\mathcal{D}}_j(k) = \{ \inneighbor{j}(k) \cap \set{D}(k) \}$.

Algorithm~\ref{alg:OpenRC} provides a synoptic view of one step of the proposed procedure for all operational modes of node $v_j$.
\begin{algorithm}[]
\caption{-- \textsc{OpenRC} iteration at agent $v_j$.}
\label{alg:OpenRC}
\begin{spacing}{1.2}
{\small
	\begin{algorithmic}[1]
         \State \noindent\textcolor{SteelBlue4}{\textbf{Arriving mode:}} $v_j\in\mathcal{J}(k)$
        \State \quad \textbf{Compute:} $x_{j}(k+1)$, $y_{j}(k+1)$, and $z_{j}(k+1)$ via \eqref{eq:rc-omasJ}

        \State \noindent\textcolor{SteelBlue4}{\textbf{Departing mode:}} $v_j\in\mathcal{D}(k)$
        \State \quad \textbf{Receive feedback:} $f_{jl}(k)$ from all $v_l \in \outneighbor{j}(k)$
        \State \quad \textbf{Send feedback:} $f_{ij}(k)$ to all $v_i \in \inneighbor{j}(k)$
        \State \quad \textbf{Assign weights:} as in \eqref{eq:c-weightstilde}.
        \State \quad \textbf{Broadcast:} $\zeta^{(x)}_j(k)$ and $\zeta^{(y)}_j(k)$ to all $v_l \in \outneighbor{j}(k)$
        \State \quad \textbf{Set:} $x_{j}(k+1)=0$, $y_{j}(k+1)=0$, and $z_{j}(k+1)=0$
        
        \State \noindent\textcolor{SteelBlue4}{\textbf{Remaining mode:}} $v_j\in\mathcal{R}(k)$
        \State \quad \textbf{Receive feedback:} $f_{jl}(k)$ from all $v_l \in \outneighbor{j}(k)$
        \State \quad \textbf{Send feedback:} $f_{ij}(k)$ to all $v_i \in \inneighbor{j}(k)$
        \State \quad \textbf{Assign weights:} as in \eqref{eq:c-weights}.
        \State \quad \textbf{Broadcast:} $\zeta^{(x)}_j(k)$ and $\zeta^{(y)}_j(k)$ to all $v_l \in \outneighbor{j}(k)$
        \State \quad \textbf{Receive:} $\zeta^{(x)}_i(k)$ and $\zeta^{(y)}_i(k)$ from all $v_i \in \inneighbor{j}(k)$
        \State \quad \textbf{Compute:} $x_{j}(k+1)$, $y_{j}(k+1)$, and $z_{j}(k+1)$ via \eqref{eq:rc-omas}

	\end{algorithmic}
}
\end{spacing}
\end{algorithm}


\begin{rem}{}
    Observe that the acknowledgement feedback does not necessarily need to be sent regularly at every time step $k$. Instead, it is only needed when a change in the activation status of the agents is made, \ie when $\grvect{\alpha}(k+1)\neq \grvect{\alpha}(k)$. 
\end{rem}
\section{Eventual Correctness Analysis}\label{sec:convergence_analysis}
Before discussing the convergence properties of the proposed algorithm, let us define the global variables that describe the evolution of the network variables. We start by collecting the local agents' variables $x_j$, $y_j$, $\widehat{x}_j$, and $\widehat{y}_j$, into the following vectors, respectively, as 
\begin{subequations}
    \begin{align*}
    \grvect{x}(k) \!&=\! \big( x_1(k), \ldots, x_n(k) \big)^{\top}\!, \; \widehat{\grvect{x}}(k) \!=\! \big( \widehat{x}_1(k), \ldots, \widehat{x}_n(k) \big)^{\top},\\
    \grvect{y}(k) \!&=\! \big( y_1(k), \ldots, y_n(k) \big)^{\top}\!, \;\; \widehat{\grvect{y}}(k) \!=\! \big( \widehat{y}_1(k), \ldots, \widehat{y}_n(k) \big)^{\top}.
    \end{align*}
\end{subequations}
Note that the values of inactive agents are also included in these variables, although (for the sake of anaysis only and without loss of generality) they are set to $0$. 

Now, we can rewrite the \textsc{OpenRC} algorithm in its vector-matrix form:
\begin{subequations}
    \begin{align*}
    \grvect{x}(k+1) &= C(k) \grvect{x}(k) + \widetilde{C}(k) \big(\grvect{x}(k)-\widehat{\grvect{x}}(k)\big)+W(k)\widehat{\grvect{x}}(k+1)\\
    \grvect{y}(k+1) &= C(k) \grvect{y}(k) + \widetilde{C}(k) \big(\grvect{y}(k)-\widehat{\grvect{y}}(k)\big)+W(k)\widehat{\grvect{y}}(k+1)
\end{align*}
\end{subequations}
where $C(k)$ and $\widetilde{C}(k)$ are matrices formed by the collection of the weights $c_{ji}$ and $\tilde{c}_{ji}$,  and
$$W(k)=\texttt{diag}(a_j(k+1)(1-a_j(k)))$$ is a diagonal matrix with diagonal entries 
$w_{jj}(k)=1$ if $v_j\in\mathcal{J}(k)$, while $w_{jj}(k)=0$, otherwise.

In the following theorem, we will show that the sum of the active agents' variables $x_j(k)$ and $y_j(k)$ are equal to the sum of their joining masses, which is preserved within the time intervals when the network is stable. With such a property at hand, we will show that active agents track the target average consensus value at all times $k \geq 0$. 

\begin{theorem}
\label{theo:preservation}
Let Assumptions~\ref{ass:1}~and~\ref{ass:2} hold true.
Then, the following holds for all $k\geq 0$:
\begin{subequations}
    \begin{align}\label{eq:induct}
    \sum_{v_j \in \set{V}(k)} x_j(k) &= \sum_{v_j \in \set{V}(k)} \widehat{x}_j(k),\\
    \sum_{v_j \in \set{V}(k)} y_j(k) &= \sum_{v_j \in \set{V}(k)} \widehat{y}_j(k).
\end{align}
\end{subequations}

\begin{proof}
    We prove the result by induction and, since the proof is the same for both $x_j(k)$ and $y_j(k)$, we only consider the first variable, \ie $x_j(k)$.  
    In order to apply our inductive argument, we observe that for $k=0$, it trivially holds 
    \begin{align}
        \sum_{v_j \in \set{V}(0)} x_j(0) = \sum_{v_j \in \set{V}(0)} \widehat{x}_j(0).
    \end{align}
    Let us now assume that \eqref{eq:induct} holds true for a given iteration $k$; we need to show that, it also holds true at $k+1$.
    In particular, noting that by removing the columns and rows corresponding to the inactive agents, we obtain a column-stochastic matrix $C(k)+\widetilde{C}(k)$, with which we have that
\begin{align}
    &\sum_{v_j \in \set{V}(k+1)} \!\!\!\! x_j(k+1)\nonumber\\ 
        &= \grvect{\alpha}^{\top}\!(k+\!1) \grvect{x}(k+1) \nonumber\\ 
        &= \grvect{\alpha}^{\top}\!(k+\!1) \! \Big(\! C(k) \grvect{x}(k) + \widetilde{C}(k)\!\big(\grvect{x}(k)-\widehat{\grvect{x}}(k)\big) +W\!(k)\widehat{\grvect{x}}(k+\!1)\!\Big)\!\! \nonumber\\
        &= \sum_{v_j \in \set{R}(k)} \!\!\!\! x_j(k) \;+ \sum_{v_j \in \set{D}(k)} \!\!\!\! \big( x_j(k)-\widehat{x}_j(k) \big) \;+ \sum_{v_j \in \set{J}(k)} \!\!\!\! \widehat{x}_j(k+1) \nonumber\\
        &= \sum_{v_j \in \set{V}(k)} \!\!\!\! \widehat{x}_j(k) \;- \sum_{v_j \in \set{D}(k)} \!\!\!\!\! \widehat{x}_j(k) \;+ \sum_{v_j \in \set{J}(k)} \!\!\!\! \widehat{x}_j(k+1) \nonumber\\
        &=  \sum_{v_j \in \set{R}(k)} \!\!\!\! \widehat{x}_j(k) \;+ \sum_{v_j \in \set{J}(k)} \!\! \widehat{x}_j(k+1) =  \!\!\!\! \sum_{v_j \in \set{V}(k+1)} \!\!\!\! \widehat{x}_j(k+1),\nonumber 
\end{align}
    where we used the fact that 
    $$\mathcal{V}(k)=\mathcal{R}(k)\cup \mathcal{D}(k),\quad \mathcal{V}(k+1)=\mathcal{R}(k)\cup\mathcal{J}(k)$$ and that $\widehat{x}_j(k+1)=\widehat{x}_j(k)$ holds for all $v_j\in \mathcal{V}(k+1)$.
    This completes our inductive proof.
\end{proof}
\end{theorem}

\begin{theorem}
    Consider that there exists a time $k^{\prime}$ where the network stabilizes, \ie no further departures and/or arrivals are allowed. Then the following holds for all $v_j \in \set{V}(k^{\prime})$ and all $k\geq k^{\prime}$:
    \begin{align}
        \lim_{k\to\infty} z_j(k) &= \lim_{k\to\infty} \frac{x_j(k)}{y_j(k)}\nonumber
        = \frac{1}{|\set{V}(k^{\prime})|} \sum_{v_l \in \set{V}(k^{\prime})} \widehat{x}_l(k').
    \end{align}
    \begin{proof}
        We start our proof by emphasizing that after time $k^{\prime}$, $\widetilde{C}(k)$ and $W(k)$ will be a zero for all $k \geq k^{\prime}$. Hence, the \textsc{OpenRC} algorithm will reduce to the ratio consensus algorithm for closed systems with initial condition ${\grvect{x}}(k^{\prime})$ and ${\grvect{y}}(k^{\prime})$. In particular, by Theorem~\ref{theo:preservation}, we have that 
        $\sum_{v_j\in\mathcal{V}(k^{\prime})}x_j(k^{\prime}) = \sum_{v_j\in\mathcal{V}(k^{\prime})}\widehat{x}_j(k^{\prime})$ and $\sum_{v_j\in\mathcal{V}(k^{\prime})}y_j(k^{\prime}) = \sum_{v_j\in\mathcal{V}(k^{\prime})}\widehat{y}_j(k^{\prime}).$
        Moreover, $C(k)$ is a column-stochastic matrix that is also primitive as long as the underlying digraph is strongly connected. Then, by utilizing the result in \cite{dominguez2010coordination}, we have that the summation of the individual iterations $\grvect{x}$ and $\grvect{y}$ is preserved and gives
        \begin{align}
            \frac{\grvect{\alpha}^{\top}(k) \grvect{x}(k)}{\grvect{\alpha}^{\top}(k) \grvect{y}(k)} \nonumber= \frac{\sum_{\ell \in \set{V}(k^{\prime})} x_\ell(k^{\prime})}{\sum_{\ell \in \set{V}(k^{\prime})} y_\ell(k^{\prime})} = \frac{\sum_{\ell \in \set{V}(k^{\prime})} \widehat{x}_\ell(k^{\prime})}{|\set{V}(k^{\prime})|},  
        \end{align}
        which implies that the ratio $z_j(k)$ converges asymptotically to the average consensus value of the agents in $\mathcal{V}(k^{\prime})$. This completes our proof.
    \end{proof}
\end{theorem}

\section{Numerical Examples}\label{sec:simulation_results}


In this section, we evaluate the performance of the proposed algorithm with numerical simulations. In what follows we illustrate the time-varying average consensus error, $e(k) := \Vert \diag\big({\grvect{\alpha}(k)}\big) \grvect{z}(k) - \mathbf{1}\bar{x}(k)\Vert_2$, where $\grvect{z}(k)=(z_1(k), \ldots, z_n(k))^{\top}$, over a network $\set{G}(k)$ which is strongly connected at every $k \geq 0$. 
To gain some insights on the performance of the algorithm, we consider $100$ initially active agents at time step $k=0$, \ie $n(0) = |\set{V}(0)|=100$, from a total of $n=|\set{V}|=150$ potentially active agents. Each initially active agent $v_j \in \set{V}(0)$ registers its joining mass $\widehat{x}_j(0)=m_j$ where $m_j$ is assumed to be chosen uniformly at random in the interval $[1,10]$, while its auxiliary variable is set to $\widehat{y}_j(0)=1$. To illustratively highlight the time-varying nature of the average consensus value in OMAS, we assume that arriving agents $v_j \in \set{J}(k)$ enter the network with a greater joining mass $\widehat{x}_j(k)=m_j$ where $m_j$ is chosen uniformly at random in the interval $[10,20]$. 

\begin{figure}[h!]
    \includegraphics[scale=0.89,left]{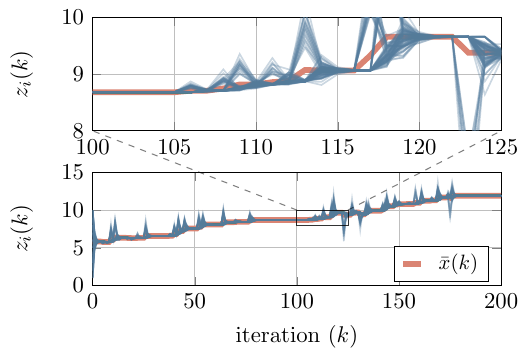}
    \caption{Evolution with time $k$, of the active agents' ratios $z_i(k)$ using the \textsc{OpenRC} algorithm.}
    \label{fig:results_ratio}
\end{figure}

In Fig.~\ref{fig:results_ratio}, we present the evolution with time of the ratio $z_j(k)$ of active agents $v_j \in \set{V}(k)$. Between the time intervals $ 1 < k \leq 80$ and $101 < k \leq 180$, the network size may increase or decrease by $1$ due to agent arrival or departure, with $10\%$ in the former interval, and $20\%$ in the latter. Between the time intervals $80 < k \leq 100$ and $180 < k \leq 200$, the network is assumed to be stable (no arrivals or departures). 
Note that, for ease of presentation, the ratio $z_j(k)$ of departing nodes is not shown in Fig.~\ref{fig:results_ratio} after their departure. The target average consensus value of the currently active agents at time $k$, \ie $\bar{x}(k)$ is shown in light red color. As it can be seen from Fig.~\ref{fig:results_ratio}, the target average consensus value $\bar{x}(k)$ is tracked by the currently active agents, while at the time instants where agents arrive to or depart from the network, their ratio $z_j(k)$ is disturbed due to the change of the network mass. 

\begin{figure}[h]
    \centering
    \includegraphics[scale=0.86,left]{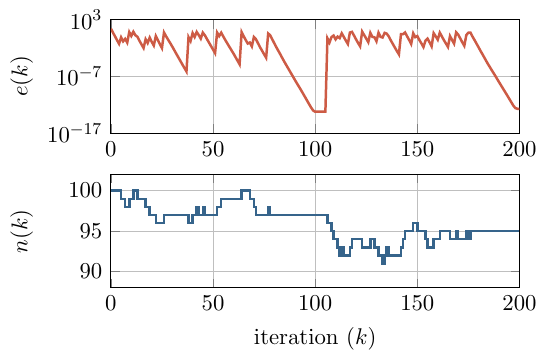}
    \caption{The average consensus error, $e(k) = \Vert \diag\big({\grvect{\alpha}(k)}\big) \grvect{z}(k) - \mathbf{1}\bar{x}(k)\Vert_2$, (upper); the number of active agents, $n(k)$, (lower), with time $k$.}
    \label{fig:results_error}
\end{figure}

In the upper plot of Fig.~\ref{fig:results_error}, the average consensus error decays in between stable network time intervals, while at the time instants where agents are arriving or departing, the average consensus error increases abruptly due to the change of the total mass of the network. However, we note that, during long enough time intervals (\ie $80 < k \leq 100$ and $180 < k \leq 200$) in which the network is stable, the average consensus error reduces up to machine precision, until the next perturbation of the network composition.

\section{Conclusions}\label{sec:conclusions}
In this paper, we proposed a novel algorithm, called Open Ratio Consensus (\textsc{OpenRC}), to address the distributed average consensus problem within Open Multi-Agent Systems (OMAS) operating over directed and possibly unbalanced networks. Using acknowledgment feedback from out-neighbors, agents show resilience to frequent changes in the composition (\ie agent arrivals and departures) of the network. Our analysis establishes mass-preservation properties as long as the out-going weights assigned by each agent sum up to 1 at every time step.
Such a property allows us to show that the active agents in the network converge to the exact average consensus value, given that after some time $k^{\prime}$ no further departures and/or arrivals occur. Our analysis is accompanied with numerical simulations that highlight the performance of the \textsc{OpenRC} algorithm in terms of minimizing the average consensus error within possibly short quiescent periods.




\printbibliography 

\end{document}